# Stable and scalable metallic phase on MoS$_2$
# using forming-gas microwave plasma

[a]Chithra H. Sharma, [a]Ananthu P. Surendran, Abin Varghese and Madhu Thalakulam[*]

School of Physics, Indian Institute of Science Education & Research Thiruvananthapuram

Kerala, India 695551

Abstract

Monolithic realization of metallic 1T and semiconducting 2H polymorphic phases makes MoS$_2$ a potential candidate for future microelectronic circuits. Though co-existence of these phases has been reported, a method for engineering a stable 1T phase in a scalable manner, compatible with the standard device fabrication schemes is yet to emerge. In addition, there are no comprehensive studies on the electrical properties of the 1T phase. In this manuscript, we demonstrate a controllable and scalable 2H to 1T phase engineering technique for MoS$_2$ using Ar + H$_2$ microwave plasma. The technique enables us to realize 1T MoS$_2$ starting from the 2H phase of arbitrary thickness and area. Our method allows lithographically defining continuous 1T regions in a 2H sample. The 1T samples withstand aging in excess of a few weeks in ambience and show a thermal stability up to 300˚C, making it suitable for standard device fabrication techniques. We conduct both two-probe and four-probe electrical transport measurements on devices with back-gated field effect transistor geometry in a temperature range of 4 K to 300 K. The 1T samples exhibit Ohmic current-voltage characteristics in all temperature ranges without any dependence to the gate voltage, a signature indicative of metallic state. The sheet resistance of our 1T MoS$_2$ sample is considerably lower than that of 2H samples while the carrier concentration of the 1T sample is few orders of magnitude higher than that of the 2H samples. In addition, our samples show negligible temperature dependence of resistance from 4 K to 300 K ruling out any hoping mediated or activated electrical transport.

---

[a] contributed equally to the work

[*] madhu@iisertvm.ac.in

# Introduction

An all two-dimensional (2D) architecture involving vertical integration of van der Waals (vW) materials has been explored as a platform for the future semiconductor technology[1-3]. Hybrid devices consisting of physically stacked layers of MoS$_2$ and other vW materials has also been explored for various device applications; MoS$_2$/Graphene interface for improved electrical contacts[4,5], MoS$_2$/h-BN hybrid systems for mobility engineering[3,5] and electrostatic confinement[6-8], MoS$_2$/WSe$_2$ PN-junction[2] have been reported. Rather than stacking, a lateral monolithic integration of regions with different electrical properties while preserving the two-dimensionality is an important ingredient for future microelectronics technology. The presence of polymorphic phases with distinct electrical properties while maintaining the layered nature makes MoS$_2$ a potential system[9-12].

Among the reported structural phases, 2H, 2H' and 3R are semiconductors, 1T' is a narrow bandgap semiconductor and 1T is metallic[9-12]. The 2H is the most widely explored phase for device applications[5,13-17]. The 1T phase has recently gained attention as a candidate for energy storage[18-20], hydrogen evolution[21-23], gas sensing[22,24] and as a low-resistance electrical contact for 2H MoS$_2$ devices[25]. The 2H phase belongs to the space group $P6_3mmc$ with a trigonal prismatic coordination between Mo and S atoms and 1T belongs to $P\bar{3}m$ space group with an octahedral coordination between the Mo and S atoms[26]. The possibility of controllably and selectively engineering metallic 1T MoS$_2$ regions starting from the semiconducting 2H MoS$_2$ provides a new route for monolithic 2D circuits.

2H to 1T phase transition happens via relative gliding of the Mo and S planes[12,27]. The transition has been achieved using alkali metal or hydrogen intercalation[11,18,25,27-30], substitutional doping by Re atom[31], annealing accompanied energetic electron-beam irradiation[12,32], plasmonic hot electrons[33] and Argon plasma[34]. Intercalation gives a mixture of 1T, 1T' and 2H phases[11,22,30,35]. The 1T phase thus obtained is reported to be thermodynamically unstable and relaxes to 1T' or 2H over time or above a temperature range of 150 °C, which is in the range of standard sample processing temperatures for device applications[26,31,35-38]. The recent report on the solution phase synthesis of 1T MoS$_2$, in large concentrations, from the 2H phase yielded only nanosheets and, is not suitable for scalable device fabrication schemes[23]. The phase conversion using Argon plasma yields predominantly the 2H phase and the concentration of 1T phase is around 40%[34]. 1T MoS$_2$ obtained by high-energy electron bombardment in a transmission electron microscope falls short in the yield and in the adaptability required for the microelectronics industry.

There is a lack of an in-depth electrical characterization of metallic 1T MoS$_2$. Existing reports confine to two-probe (2P) transport measurements on polymorphic samples[11,25,34].

A four-probe (4P) electrical characterization is essential since 2P measurements are influenced by the behaviour of contacts. A linear current-voltage (I-V) characteristics, void of gate-voltage dependence, down to the cryogenic temperatures is necessary in establishing a metallic nature. The 1T samples obtained by Argon plasma[34] treatment have shown response to gate voltages, atypical of a metallic state and, in contrast to those reported elsewhere[11,25]. Temperature dependent transport measurements are reported only down to 100 K, however, the sample shows a large increase in the resistance as the temperature is lowered[11].

In this manuscript, we demonstrate a method to create large area 1T $MoS_2$ sheets, from bulk exfoliated 2H $MoS_2$ samples with arbitrary thickness and area, in a controllable and scalable manner. Our process involves treating mechanically exfoliated samples with high-power forming-gas microwave plasma which results in a layer-by-layer thinning accompanied by a structural phase conversion from 2H to 1T. The presence of plasma etching helps us to realize large-area few-layer 1T $MoS_2$ samples starting from large-area thicker exfoliated samples. We show that our technique can be used to selectively engineer the 1T phase on 2H $MoS_2$ samples with the help of standard lithography techniques. We perform an in-depth structural characterization using high-resolution transmission electron microscopy (HR-TEM) and Raman spectroscopy. Unlike the intercalation route, our process is faster and we do not find signatures of 1T' or 1T" phases. Our process yields extended 1T regions with an areal coverage in excess of 70 % over the 2H phase. The 1T samples show a temporal stability in excess of a few weeks and a thermal stability up to 300° C in ambience. We conduct 2P transport studies on lithographically defined 2H and 1T regions on the same sample for a direct comparison of electrical properties. We also perform 4P electrical transport studies on a few layer 1T $MoS_2$ sample. The contacts on the 1T phase show a clear Ohmic behaviour at all temperatures from 300 K down to 4 K and the transport show little response to the gate voltage, indicative of a metallic phase. We also observe negligible temperature coefficient of resistance down to 4 K for the 1T phase unlike other reports on 1T $MoS_2$[11], ruling out hoping-mediated or activated transport in our samples.

**Experimental methods**

The $MoS_2$ samples with various thickness are mechanically exfoliated from bulk crystals and transferred onto Silicon substrates hosting a 300 nm thermally grown silicon-oxide layer using the PDMS dry stamping technique[39]. The thickness of the flakes is estimated using the red-channel optical contrast method[40]. These samples are treated with forming gas (10% $H_2$+90% Ar) microwave plasma with 40% input power to the magnetron. The plasma treatment results in a layer-by-layer etching of the sample accompanied by a structural phase transformation from the 2H to the 1T phase. Technical details of the microwave plasma reactor is given elsewhere[40]. Lower microwave power levels result only in a layer-by-layer etching and do not yield any phase change.

The HR-TEM analysis of the samples are carried out using FEI Tecnai FEG30 E Spirit transmission electron microscope with an accelerating voltage of 200 kV. The samples are transferred onto a TEM grid from the Si/SiO$_2$ substrate using the PMMA transfer technique[41]. The Raman and photoluminescence (PL) spectra are recorded using a Horiba XploRA PLUS Raman microscope using a laser of wavelength 532 nm with a power of 1.5 mW.

For selective area phase conversion, we exploit a lithographically defined Aluminium mask during plasma treatment. Post plasma treatment, the mask is removed using 0.1 N NaOH solution. Field effect transistors are fabricated for exploring the electrical properties of the material; MoS$_2$ film acts as the channel, while the highly doped silicon substrate and the silicon oxide layer serves as the back-gate and the gate-dielectric respectively. The source and the drain contacts are defined using standard electron-beam or photo lithography followed by Cr/Au metallization. The transport measurements are performed in high vacuum (<10$^{-6}$ mbar) dark environment in a closed-cycle cryostat.

## Results & Discussion

**TEM analysis**

The crystallinity of the plasma treated samples is examined using HR-TEM. Fig. 1(a) shows HR-TEM image of a representative plasma treated few-layer MoS$_2$ sample. The regions shaded in purple, green and brown represent the 2H, 1T and, an intermediate state between the 2H and the 1T phases respectively. From the HR-TEM analysis conducted over many samples we estimate an areal coverage in excess of 70 % for the 1T phase opposed to that of the 2H phase. We also note that the HR-TEM images do not show signatures of other structural phases such as 1T' or 1T". The selected area electron diffraction (SAED) pattern shown in the top-right inset exhibit sharp diffraction spots, inferring good crystallinity of our samples. HR-TEM images of two more samples showing extended 1T regions are shown in supporting information S1.

Fig. 1(b) shows a magnified image of the 2H region while Fig. 1 (c) shows that of the 1T region obtained from the same sample as in Fig. 1 (a). The lower panels in Fig. 1(b) and (c) shows the intensity line-profiles obtained along the directions indicated by the dashed-lines in the respective panels. The arrangements of Mo and S atoms for the 2H and 1T phases are shown in the overlaid diagrams in Fig. 1 (b) and (c) respectively. In the 2H phase the S-Mo-S atoms are arranged in an A-B-A stacking fashion along the c - axis.  Mo atoms appear brighter in intensity compared to the S atoms in the TEM images owing to its higher atomic number[42]. Due to this reason, the intensity for the S-peaks is very weak for monolayer MoS$_2$. For a few-layer 2H MoS2, the position of S atoms in one layer coincides with that of the Mo atoms in the adjacent layer. This gives an appreciable intensity for the peaks corresponding to the S sites[43], as evident from the intensity profile shown in the bottom panel of Fig. 1 (b). We extract a nearest Mo-Mo separation of 3.19 (+/- 0.08) A⁰ from the HR-TEM images.

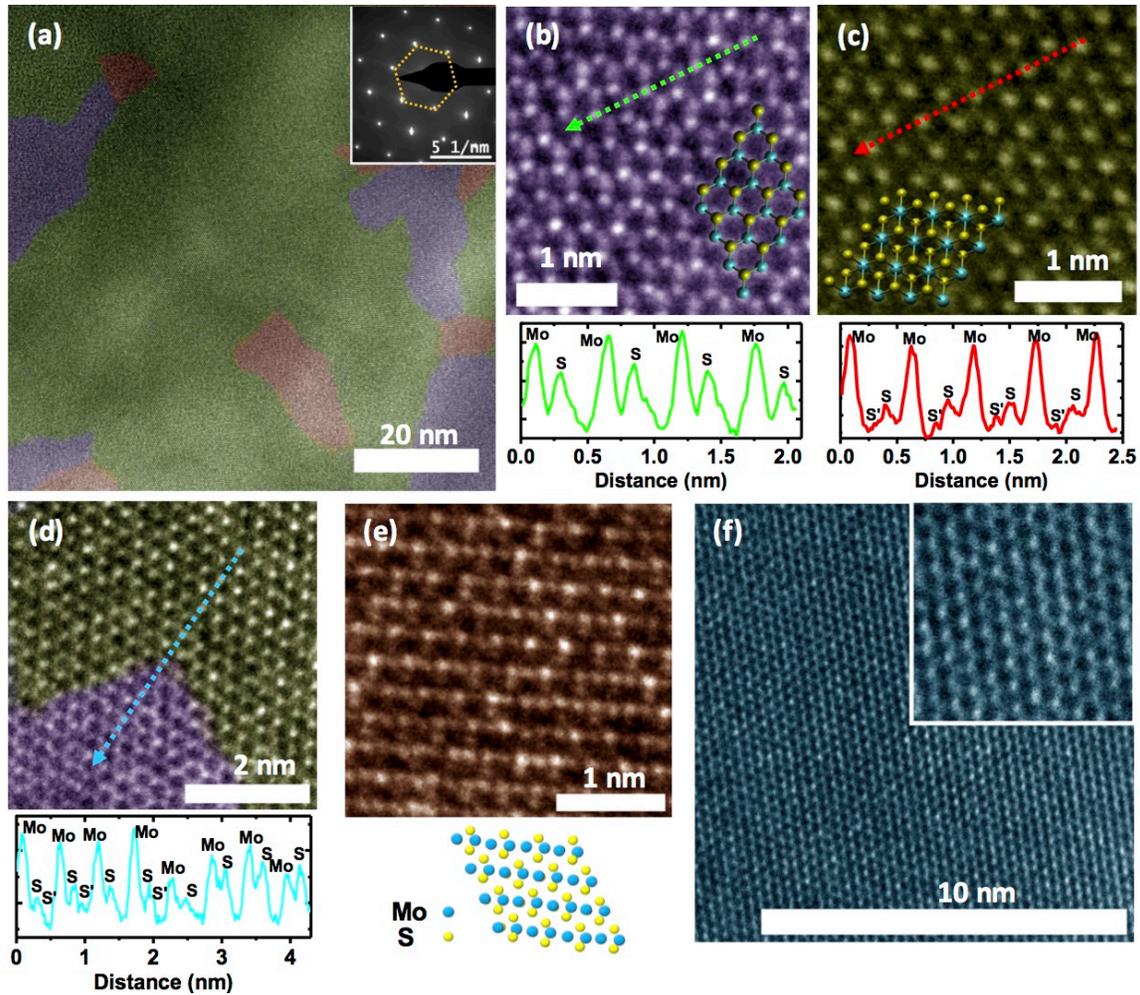

Figure 1: (a) HR-TEM image of a plasma treated MoS$_2$ sample showing 2H-phase (purple), 1T –phase (green) and an intermediate state (brown). The inset shows SAED pattern from the sample showing sharp diffraction spots. (b) & (c) Post-plasma treatment HR-TEM images of the 2H & the 1T regions respectively. The corresponding atomic structure of the 2H and the 1T phase overlaid, cyan spheres represent Mo sites and yellow spheres represent S sites. The bottom panels show intensity profile taken along the directions represented by the dashed-line in the corresponding TEM image. (d) HR-TEM image from a region where both 2H and 1T regions intersect. The bottom panel shows the intensity profile taken along the dashed-line in the TEM image. (e) HR-TEM image showing an intermediate state between the 2H and the 1T phases. Bottom panel shows representative atomic arrangement. (f) HR-TEM image from a pristine MoS$_2$ sample showing the 2H phase. Inset shows a magnified view, 3 nm x 3 nm in area.

In the 1T phase, the Mo atom is octahedrally coordinated with six S atoms with the S-Mo-S in an ABC stacking fashion. In this case, atoms in one layer align with the corresponding atoms in the adjacent layer. This arrangement makes the intensity of the peaks corresponding to the S atoms in 1T phase much weaker compared to that of the 2H phase in the TEM images. As seen in the bottom panel of Fig. 1 (c) there are two small peaks corresponding to the Sulphur atoms in the top and bottom plane (S and S') while the peaks for Mo atoms are clearly visible. For the 1T phase the Mo atoms form a hexagonal lattice as shown in Fig. 1(c). We obtain a mean nearest Mo-Mo separation of 3.19 (+/- 0.04) A⁰ from line-profiles taken along different directions in the (001) plane [Supporting Information S2]. These values are in good agreement with the reported ones[44]. We rule out the presence of any 1T' or 1T" phases where Mo-Mo separations are unequal in different directions, as observed elsewhere[45].

Fig. 1 (d) shows a magnified HR-TEM image of a region where the 2H and 1T phases intersect. The lower panel shows intensity line-profile taken along the direction represented by the dashed-line in the main panel of Fig. 1 (d). 2H and the 1T regions exhibit intensity profiles similar to Fig. 1 (b) and (c) respectively. We also note that the relative intensity of the Mo peaks in the 1T region is higher than that of the 2H region; possibly due to the difference in the alignment of Mo atoms, corresponding to different layers, along the c-axis.

The transformation between the 2H and the 1T phases involves an intra-layer S plane gliding[12,27]. For a few layer MoS₂ sample, transition between the 2H and 1T phases also require an Mo-plane gliding. In support of this we observe in a few of our HR-TEM images, an intermediate atomic arrangement between the 2H and the 1T phases, as shown in Fig. 1(e). The visible stripe-like patterns in Fig. 1 (e) is due to the rearrangement of Mo and S atoms during the transformation[28]. A possible atomic arrangement is shown in the inset to Fig. 1(e).

We also examine HR-TEM image of a pristine 2H MoS₂ sample taken with the same exposure parameters as those for the 1T samples. Fig. 1 (f) shows the HR-TEM image and the inset shows a magnified view of the Mo and S atomic arrangements depicting the 2H phase. The images did not show presence of any other structural phases confirming the absence of electron-beam induced 2H to 1T phase transition in our samples.

**Raman & PL studies**

We perform Raman scattering and PL studies on plasma treated samples. These samples consist of regions with different thickness starting from a few nanometres to a few tens of nanometres prior to the plasma treatment. Fig. 2 (a) shows optical images of a representative sample where the top panel shows the pristine exfoliated sample while the bottom panel shows the image after a 7.5 minutes of plasma treatment. We have conducted Raman scattering studies on all the regions. Here we focus on regions labelled I and II. Fig. 2(b) [Fig. 2 (c)] shows the Raman spectra from region I [region II]. The black(red) traces in both Fig. 2(b) and (c) represent Raman spectra taken before(after) the plasma treatment. Samples post-plasma treatment (red traces) show clear $J_1$ and $J_2$ vibrational modes corresponding to the 1T phase (The Raman scattering studies conducted on other regions are shown in Supporting Information S3).

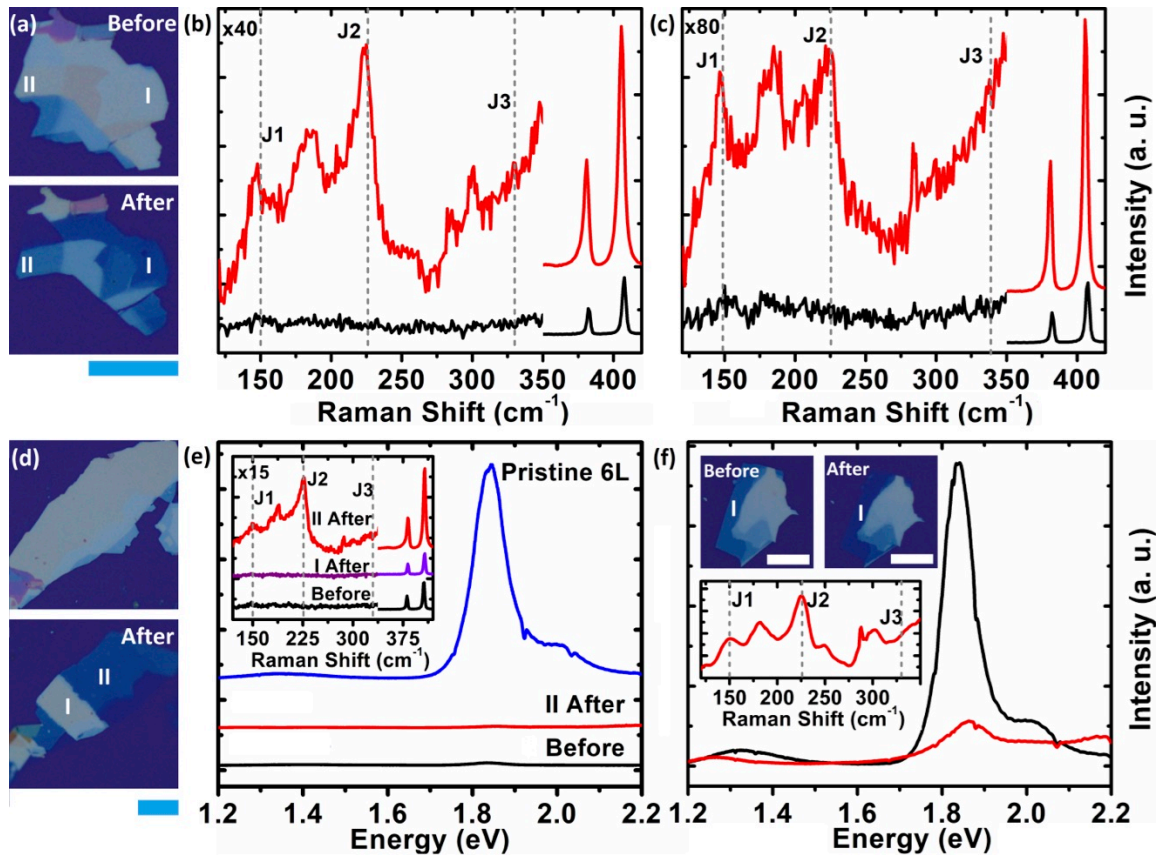

Figure 2: (a) Optical images of a representative sample before (top) and after (bottom) plasma treatment. (b) & (c) Raman spectra of the, sample shown in (a), taken from region I & II respectively. Black-traces (Red-traces) correspond to the Raman spectra taken before (after) plasma treatment. Positions of $J_1$, $J_2$ and $J_3$ peaks are marked. The amplitude of the spectra is multiplied by the factor indicated in the plot for Raman shifts below 350 cm$^{-1}$. (d) Optical images of a pristine sample (top) and after plasma treatment (bottom). The central region marked I is masked during plasma treatment. (e) PL spectra taken from region II, of the sample shown in (d), before (black) and after (red, ~ 6 layers) plasma treatment. The blue-trace shows PL from a pristine six-layer sample for comparison. The inset shows Raman spectra from the region I (magenta) and II (red), of the sample shown in (d). The black-trace shows Raman spectra of the sample before plasma treatment. The y-axis is multiplied by the factor indicated in the plot for Raman shifts below 350 cm$^{-1}$. (f) The PL spectra from region I of a sample before (~ 6 layers, black-trace) and after (~ 4 layers, red-trace) plasma treatment. The top-insets show optical images of the sample before and after plasma treatment. The bottom inset shows Raman spectra, post plasma treatment, showing the $J_1$, $J_2$ and $J_3$ peaks. All the scale bars correspond to 20 μm.

To demonstrate the controllability and scalability of the process we use the Aluminium masking technique to selectively phase engineer the sample. Fig. 2 (d) shows optical images of the sample before (top) and after (bottom) the plasma treatment. The centre region (labelled I) on the bottom panel of Fig. 2 (d) is masked and the remaining area is treated with the plasma and etched down to a thickness of ~ 6 layers of $MoS_2$. Fig. 2 (e) shows the PL spectra of the sample before (black) and PL spectra from region II after (red) the plasma treatment. The sample is in excess of ~ 50 nm in thickness prior to plasma treatment and exhibits only a weak excitonic peak (black trace). The plasma treated region (II), reduced to 6 layers in thickness, does not exhibit any PL (red trace) in contrast to a pristine six-layer 2H $MoS_2$ shown in blue-trace[46]. Raman spectra of the sample before and after the plasma treatment are shown in the inset. Only region II, exposed to the plasma, develops the $J_1$ and the $J_2$ peaks (red trace) while the Raman spectra of region I post plasma treatment (purple trace) is akin to that of the sample prior to the plasma treatment. The quenching of the PL spectra, post plasma treatment, as a result of a semiconducting to metallic phase transition is shown in Fig. 2 (f). The optical images of the sample before and after plasma treatment is shown in the insets. The black-trace shows the PL spectra obtained from region I (~ 6 layers) and the red trace shows the PL spectra from the same region after the plasma treatment (~ 4 layers). While the PL spectra undergo a substantial quenching, the Raman spectra, shown in the inset, develop the characteristics $J_1$ and $J_2$ peaks as a result of the plasma treatment.

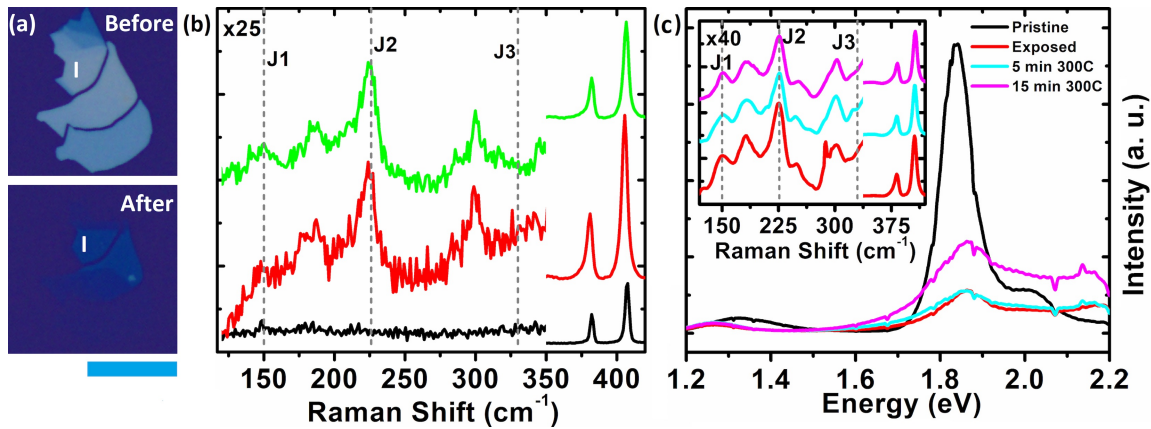

Figure 3: (a) Optical images of the sample before (top) and after (bottom) plasma treatment, scale bar is 20 μm. (b) Raman spectra of the sample, shown in (a), before (black) and after (red) plasma treatment with positions of the $J_1$, $J_2$, and $J_3$ peaks marked. The green trace in (b) shows the Raman spectra taken after 27 days from plasma treatment. (c) PL spectra from region I of the sample shown in Fig. 2(f) before (black), after (red) plasma treatment and, subsequently annealing the sample at 300°C for 5 minutes (cyan) and 15 minutes (magenta). The inset shows the Raman spectra for the same with corresponding colours.

Post plasma treatment, our samples exhibit clear $J_1$ and $J_2$ peaks while the $J_3$ peak is very weak in intensity. In addition, we observe emergence of a peak at 180 cm$^{-1}$ which was not reported in the past. The $J_1$ and the $J_3$ peaks are predicted to be much lower in intensity compared to the $J_2$ peak[47]. The relative intensities and the peak positions vary from sample to sample and process to process[20,29,32,34,48]. 1T $MoS_2$ prepared by chemical routes exhibit a weaker $J_3$[29,48]. 1T $MoS_2$ prepared by physical routes exhibit strong $J_2$ while the $J_3$ is very weak. Both the $J_1$ and $J_2$ peaks are clearly visible on electron-beam irradiated and Argon RF plasma treated samples while $J_3$ is not well formed[32,34].

Now we discuss the stability of the phase engineered 1T samples. Fig. 3 (a) shows the optical images of the sample before and after plasma treatment. Raman spectra from the region I of the sample before (black-trace) and after (red-trace) the plasma treatment is shown in Fig. 3 (b). The green-trace shows Raman spectra obtained from the same region after keeping the sample for 27 days at ambient temperature and pressure. Both the red-trace and the green-trace show $J_1$, $J_2$ and $J_3$ peaks with similar shape and intensity suggesting good temporal stability of our samples. We also note here that the HR-TEM images taken after 30 days of plasma treatment also show rich concentration of the 1T phase suggesting good temporal stability of our samples.

Now we explore the stability of the 1T samples as a function of annealing temperature. Fig. 3 (c) shows the PL spectra taken from the region labelled I of the sample shown in Fig. 2 (f) post plasma treatment at room temperature (red-trace), annealed at 300° C for 5 min (cyan-trace) and 15 min (magenta-trace). The black trace shows the PL spectra of the same region, ~ 6 layers in thickness, prior to the plasma treatment. The inset shows Raman spectra of region I at room temperature (red), annealed at 300° C for 5 min (cyan) and for 15 min (magenta). We infer from the PL and Raman spectra evolution that our 1T samples are stable up to 300° C in temperature, which is well above the range of temperatures for standard device fabrication processes. As evident from the HR-TEM images our samples also contains regions in 2H phase which might be providing stability to the 1T phase[32].

**Transport studies**

Electrical transport studies are conducted on two kinds of devices. (1) On a device where a 1T region is lithographically defined on a pristine 2H $MoS_2$, for a direct comparison of electrical properties of both phases on the same sample. (2) On a fully phase engineered 1T MoS$_2$ device.

Fig. 4 (a) shows the optical image of the sample on which the central region enclosed between the dashed- lines (labelled 2H) is covered using a lithographically defined Al mask, and the flanked regions labelled 1T are exposed to the plasma and converted to the 1T phase. Post plasma treatment the mask is removed using NaOH (0.1 N) and Cr/Au source and drain contacts are fabricated onto both the 1T and the 2H regions. The Raman spectra of the 1T region shown in the Supporting Information S4 exhibit the signature peaks, the J$_1$, J$_2$ and J$_3$, of 1T phase. Fig. 4 (b) shows the 2P I-V characteristics of the 2H region at 300 K (red trace), 77 K (green trace) and 4 K (blue trace). The I-V characteristics of the 2H-region exhibit a Schottky behaviour at all temperatures and the span of the non-linearity increases as the temperature is lowered down to 4K. We note that Cr/Au contacted MoS$_2$ generally exhibit non-linear I-V characteristics[13,40], which has been verified by us independently (data not shown). The inset to Fig. 4 (b) shows the 2P conductance of the 2H region as a function of the back-gate voltage, $V_{bg}$, at 300 K (red trace) and 4K (blue trace). The device exhibits clear n-type behaviour with a field effect mobility of 16.4 cm$^2$/V-s at 300 K and 84 cm$^2$/V-s at 4 K, which are in the range of typically observed mobility values for an uncapped, back-gated 2H MoS$_2$ FET[49]. In contrast, the I-V characteristics of the plasma treated region 1T, shown in Fig. 2(c), exhibits excellent Ohmic behaviour at all temperatures down to 4K. The inset shows the conductance of the 1T region as a function of $V_{bg}$. The device shows little change in conductance as $V_{bg}$ is varied in a large voltage range of -20 to 40 V at 300 K (red trace) and -10 to 20V at 4K (blue trace). We also note that the 2P resistance of the 1T region shows only a small change (~ 12 Ohms) as the sample was cooled down to 4K from 300 K while that of the 2H region shows a large variation in excess of three orders in magnitude. We have conducted 2P transport measurements on a monolayer phase engineered 1T MoS$_2$ sample and verified the linear I-V characteristics and the absence of back-gate voltage dependence on conductance independently [Supporting Information S5].

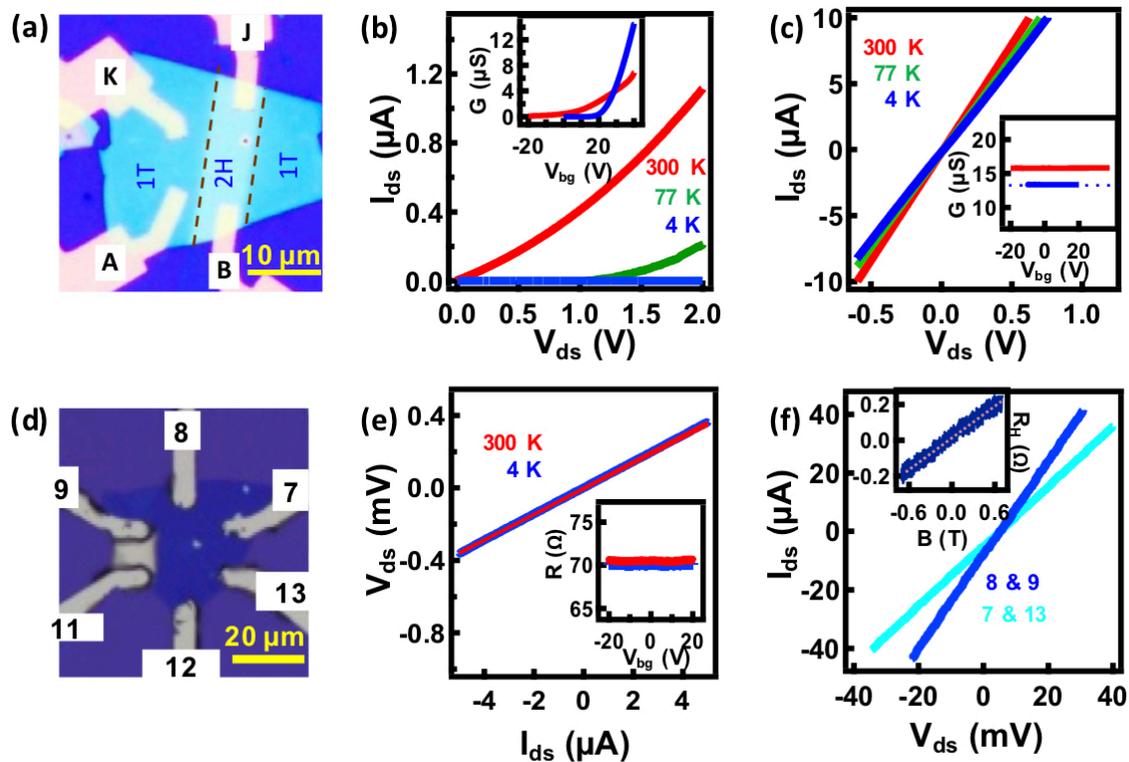

Figure 4: (a) Optical image of a selectively phase engineered sample. The region labelled 2H is masked and the regions labelled 1T are exposed during plasma treatment. (b) & (c) I-V curves from 2H & 1T regions respectively; 4 K (blue), 77 K (green) and 300 K (red). Insets show conductance of the corresponding regions as a function of V$_{bg}$ at 4 K (blue) and 300 K (red). (d) Optical image of a ~ 8 nm thick phase engineered 1T MoS$_2$ sample with photo-lithographed electrical contacts, scale bar is 20 μm. (e) 4P I-V characteristics of the sample at 300 K (red) and 4 K (blue), shown in (d); probes labelled 8 & 9 are used for current sourcing and probes labelled 7 & 13 are used for voltage sensing. The Inset shows the resistance vs V$_{bg}$ at 300 K (red) and 4 K (blue). (f) 2P I-V characteristics of the sample, shown in (d), taken across the probes 8 & 9 (blue) and 7 & 13 (cyan) at 4K, showing a clear Ohmic behaviour. The inset shows Hall resistance of the sample at 4 K.

To exclude any contribution from the contact resistance to the electrical characteristics we conduct 4P transport measurements on an ~ 8 nm thick 1T phase engineered sample. The optical image of the device is shown in Fig. 4(d). The 4P I-V characteristics of the sample at 300 K (red trace) and 4 K (blue trace) are shown in Fig. 4 (e). The I-V characteristics show clear Ohmic behaviour down to 4 K. We observe a feeble change in the resistance as the device is cooled down to 4K; we extract a temperature coefficient of resistance, $\alpha = -1.1057 \times 10^{-4} K^{-1}$ using the equation $R_T = R_{300K}[1 + \alpha \Delta T]$. The inset shows the four-probe conductance of the device as a function of $V_{bg}$. Both at 300 K (red trace) and at 4 K (blue trace) the conductance of the device does not show any response to $V_{bg}$ for a large range of voltage. Fig. 4 (e) shows 2P I-V characteristics of the voltage (8 & 9) and current (7 & 13) probes of the device at 4 K, both exhibiting excellent linearity. From the 4P resistance we extract a sheet resistance = 108 Ω/□. A carrier concentration of ~ 2.3 X 10$^{15}$ cm$^{-2}$ is extracted from the Hall resistance shown in the inset to Fig. 4 (e). We note here that the carrier concentration obtained for our 1T samples are higher by two orders of magnitude compared to that of back-gated 2H MoS$_2$ samples[13] and higher approximately by an order compared to that of the ionic-liquid-gated 2H MoS$_2$ samples[50].

## Conclusions & discussions

In this manuscript, we demonstrated a controllable and scalable 2H to 1T phase conversion technique for MoS$_2$. The process involves treating exfoliated 2H MoS$_2$ of arbitrary thickness with forming-gas microwave plasma. We performed an in-depth structural analysis using HR-TEM and Raman microscopy. Our processed samples consist of 70% of 1T phase. We did not find presence of other commonly observed phases such as 1T'or 1T". The HR-TEM images showed clear signatures of Mo and S atomic plane gliding revealing the mechanism of the phase transition. We observed the evolution of the signature Raman peaks of 1T MoS$_2$ accompanied by quenching of the PL on plasma treatment, indicative of a metallic phase formation. Our 1T samples withstood aging for more than a month and also showed a thermal stability up to 300˚C, making it suitable for standard device fabrication techniques. We demonstrated lateral monolithic integration of metallic 1T and semiconducting 2H phases with the help of standard lithography techniques. We have conducted extensive transport characterization of our 1T samples from 300 K down to 4 K. Both the 2P and the 4P I-V characteristics showed excellent linearity down to 4K and did not exhibit any response to the back-gate for a large span of voltage. Our 1T samples showed carrier concentration a few orders higher and the resistance considerably lower than that of the 2H samples. A linear I-V characteristic, without gate-voltage dependence suggests the presence of a metallic state and Schottky-barrier-free source and the drain contacts[51,52]. The feeble temperature dependence shown by the 1T samples from 300 K down to 4K also negates any barrier formation at the source and drain contacts[53]. The negligible temperature dependence also rules out any hoping mediated and activated transport in our system and suggests that the samples consists of extended 1T regions as evident from the HR-TEM analysis; transport on polymorphic MoS$_2$ had shown strong temperature dependence due to hopping transport between 1T islands[11]. The weak temperature dependence could also be due to the phonon-decoupling effect as observed elsewhere[54]. Scalability and compatibility with planar device fabrication schemes, high yield, and stability of the samples can make this process a promising tool for 2D microelectronics industry.

## Acknowledgement

The authors acknowledge IISER Thiruvananthapuram for the infrastructure and experimental facilities and, D. D. Sarma for useful discussions. M.T. acknowledges the funding received from DST-SERB extramural program (SB/S2/CMP-008/2014). C.H.S. acknowledges CSIR and A.P.S. acknowledges INSPIRE for the fellowship.

# Supporting Information

# Stable and scalable metallic phase on MoS$_2$ using forming-gas microwave plasma

Chithra H. Sharma, Ananthu P. Surendran, Abin Varghese and Madhu Thalakulam[*]

School of Physics, Indian Institute of Science Education & Research Thiruvananthapuram
Kerala, India 695551
[*] madhu@iisertvm.ac.in

## S1: HR-TEM showing 2H and 1T regions

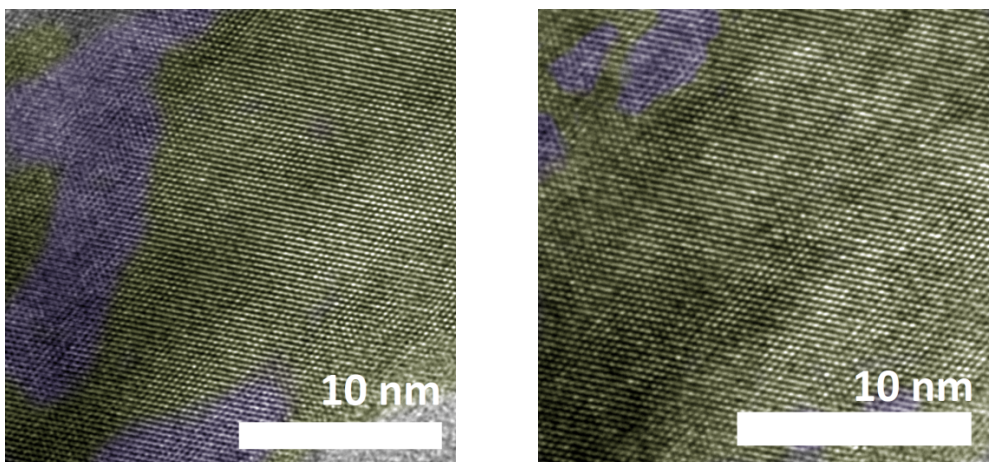

Representative HR-TEM images of plasma treated samples showing 1T (green) and 2H (purple) phases.

## S2: Line profile showing Mo-Mo distance

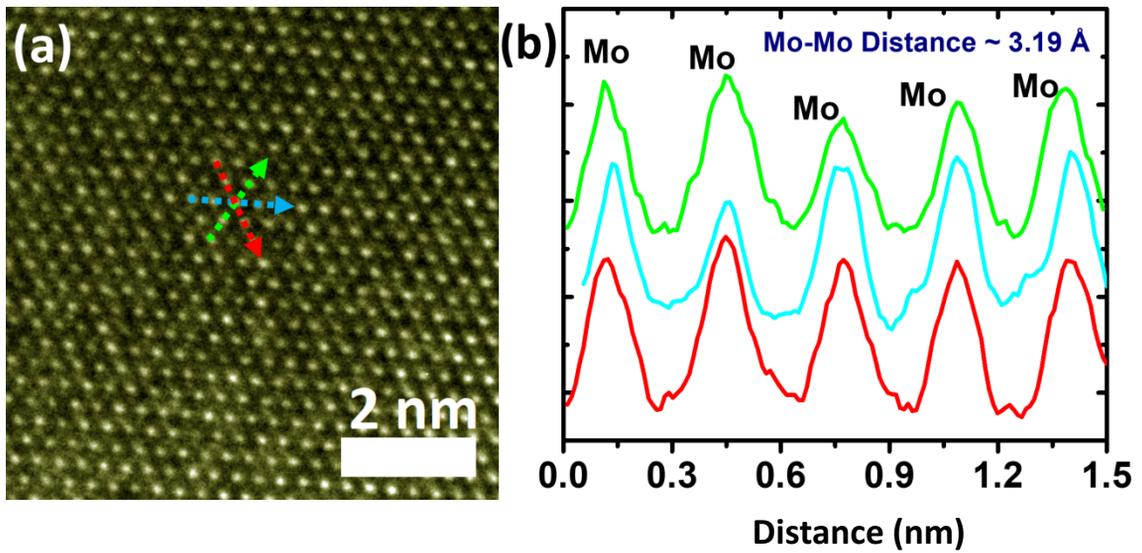

The Mo-Mo bond distances in three different directions showing constant value indicating that it is not the 1T' phase.

## S3: Raman showing J1, J2 peaks post-plasma treatment

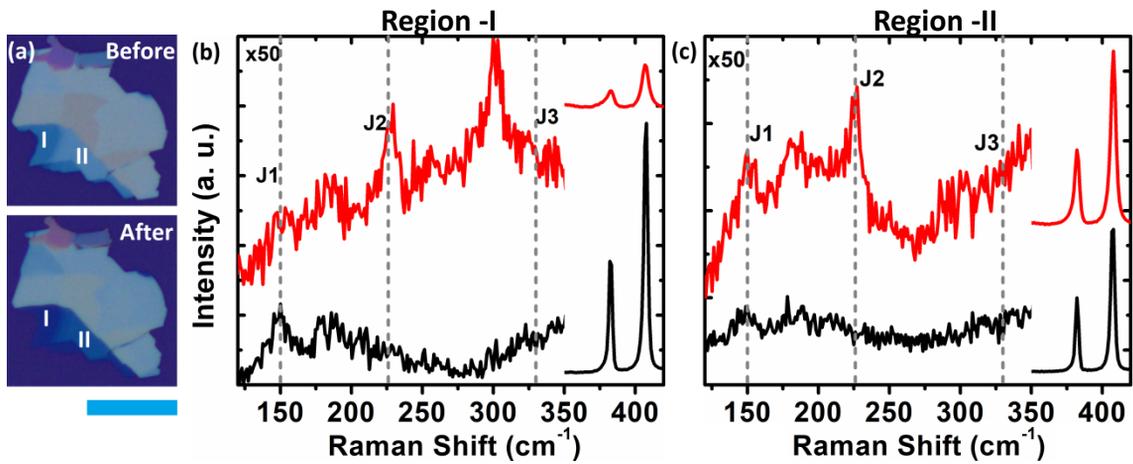

Raman from region I and II before (black) and after (red) plasma treatment show clear peaks corresponding to the $J_1$ and $J_2$ vibrational modes corresponding to the 1T phase. The region I post plasma corresponds to a monolayer.

## S4: Raman spectra from the 1T region of sample in Fig. 4 (a)

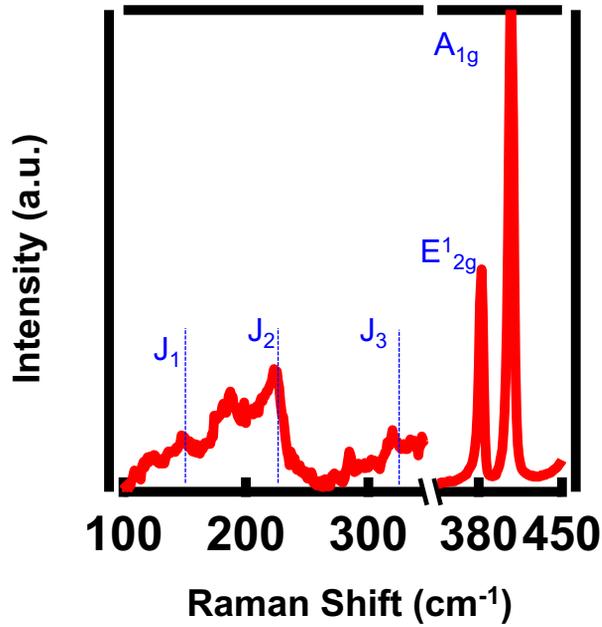

The Raman spectra from the 1T region of the sample in Fig. 4 (a) showing $J_1$, $J_2$, $J_3$, $E^1_{2g}$ and $A_{1g}$ peak positions

## S5: Gate response and IV from a plasma treated sample

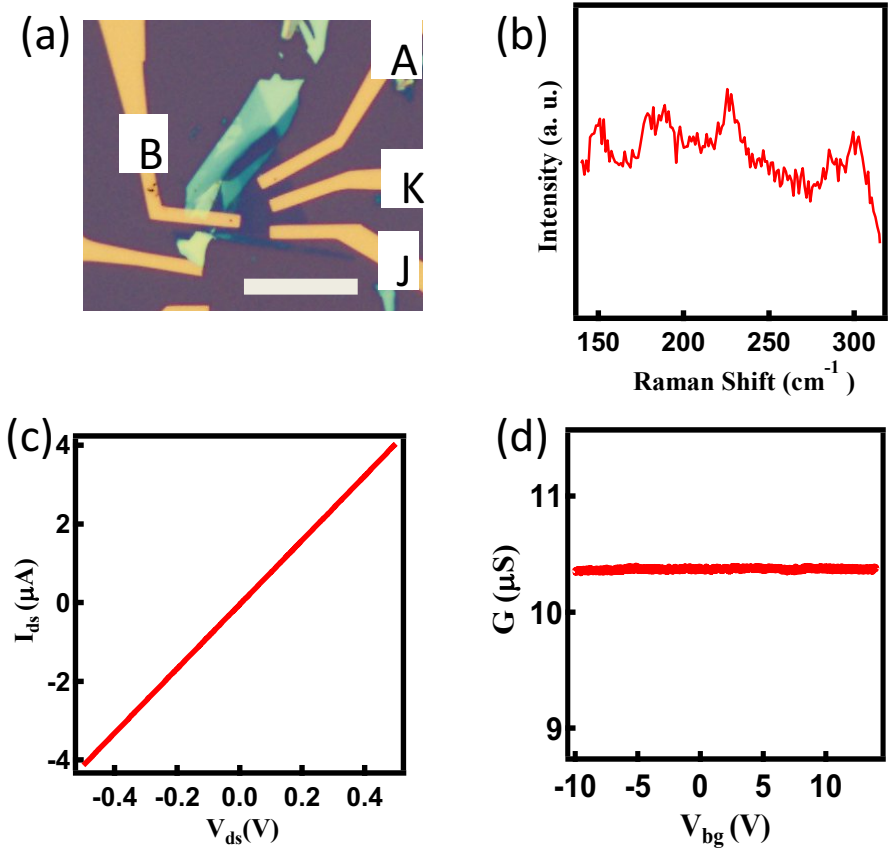

Image of the device on plasma treated sample on the top. The dependence of conductance on $V_{bg}$ at RT with Raman showing clear peaks corresponding to the $J_1$ and $J_2$ vibrational modes corresponding to the 1T phase (left inset) and I-V characteristics (right inset)